**Comment on „Role of inelastic tunneling through the insulating barrier in scanning-tunneling-microscope experiments on cuprate superconductors" [1].**

In a recent letter [1] Pilgram, Rice and Sigrist (PRS) propose to perform oxygen isotope dependent conductivity measurements along the c-axis of the cuprate superconductor BSSCO. In Raman isotope data for vibrations along the c-axis from 1988 [2], and in site selective susceptibility experiments determining $T_c$, no isotope effects were observed for oxygen ions in the insulating layer [3]. Based on these earlier not by PRS quoted experiments, we argue that the suggestion made in Ref. 1 is not worth to be undertaken, and if it is carried through, no effect will be seen. From the meanwhile 12 years old site selective data [3] the observed isotope effects stem to nearly 100% from oxygen ions in the $CuO_2$-planes. These results evidence polaronic behavior in these planes [4].

In the letter [1] previously published STM data [5] are reanalyzed in terms of inelastic tunneling through the apical oxygen ions. A known polaron model with coupling of an oxygen p-orbital to a dispersionless c-axis phonon is used under simplifying assumptions. This effort yields reasonable agreement with the recorded differential conductance data [5]. The authors then concentrate on the apical oxygen ion, and show within their model that the isotope effect is present as well. However, based on the non-detectable oxygen isotope effect for vibrations along the c-axis of the material [2] as well as a non observable isotope effect on $T_c$ for the apical oxygen ion [3] it is unlikely to envisage that a site selective differential conductivity experiment will detect any effect. If nevertheless such an experiment is carried out without success, this does not imply the absence of polarons. These, however, are formed in the $CuO_2$-planes. Susceptibility, EXAFS, photoemission, EPR, X-ray absorption, pulse probe data demonstrate their presence [4].

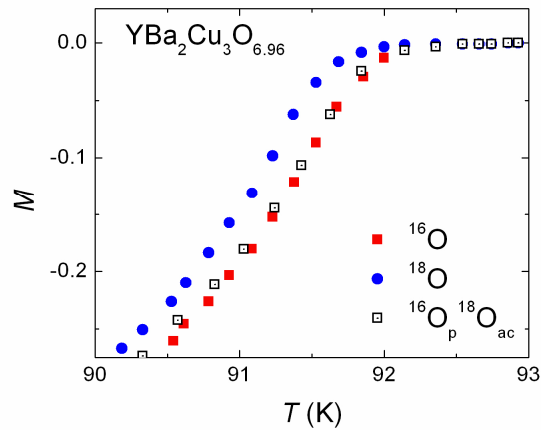

**Figure 1** Magnetization curves near $T_c$ showing the site selective oxygen-isotope effect in optimally doped $YBa_2Cu_3O_{6.96}$. $^{16}O$ and $^{18}O$ indicate fully oxygen exchanged samples. $^{16}O_p^{18}O_{ac}$ denotes a site selective substituted sample in which the planar (p) oxygen sites are occupied by $^{16}O$, whereas the apex (a) and chain (c) sites are occupied by $^{18}O$. After Ref. [3]; also confirmed for lower doping in Ref. [6].

The role of the apical oxygen ions for superconductivity has been addressed early on [7, 8, 9] using similar Hamiltonians as employed in [1]. In addition, shortly after the discovery of HTSC in cuprates [10], it has been suggested that anharmonic electron-lattice coupling of the apex oxygen ion could be relevant to superconductivity [11] since it plays a key role for the chain-$CuO_2$-plane potential which exhibits a double minimum potential [8, 12]. Subsequently, this possibility as well as the role of planar oxygen ions were tested experimentally by initiating an intense program on isotope effects in cuprates, including those on the pseudo-gap temperature, the penetration depth, and the afore mentioned site selective isotope experiments on $T_c$ [3]. In Fig. 1 the main result is shown evidencing that <u>no</u> substantial oxygen isotope effect stems from the apex position [3, 8].

Besides of the above failure of PRS to relate their work to previous results, we would like to point out that the isotope effect reported by STM methods [5], is marginal only, since the $T_c$ values of the substituted and the unchanged samples differ substantially and in the opposite way than expected. This implies that <u>different doping levels are present in the two samples</u> [13]. This is in contrast to experimental results (Fig. 1) which have been checked carefully by back exchange.

In summary, based on existing experiments and theory, no oxygen isotope effects for ions in the apical positions should be observed in differential conductivity experiments as predicted by PRS [1]. However, the known zero effects do <u>not</u> exclude polaronic states in cuprate HTS. They manifest themselves <u>in</u> the all important $CuO_2$-planes. The planar oxygen ions carry a substantial isotope effect (Fig. 1). These are not in the tunneling path of Ref. 5 and may therefore give only weak second order contributions to the c-axis tunneling current. Consequently it is unlikely that Lee et al. [5] are able to detect the glue boson.

A. R. Bishop, S. D. Conradson
Los Alamos National Laboratory
Los Alamos, NM87545, USA

A. Bussmann-Holder, O. Dolgov
Max-Planck-Institut für Festkörperforschung
Heisenbergstr. 1, D-70569 Stuttgart, Germany

H. Keller, R. Khasanov, K. A. Müller
Physik Institut der Universität Zürich
Winterthurerstr. 190, CH-8057 Zürich, Switzerland

V. Z. Kresin
Lawrence Berkeley Laboratory
1 Cyclotron Road, Berkeley CA94720, USA

H. Kamimura
Department of Applied Physics, Faculty of Science
Tokyo University of Science, 1-3 Kagurazaka, Shinjuku-ku, Tokyo, Japan 162-8601

D. Mihailovic
Complex Matter Department, Josef Stefan Institute
Jamova 39, SI-1000 Ljubljana, Slovenia